\documentclass[aps,prc,twocolumn,superscriptaddress]{revtex4-1}

\usepackage{color}
\usepackage[T1]{fontenc}
\usepackage{selinput}
\usepackage{changes}

\usepackage{graphicx}
\usepackage{dcolumn}
\usepackage{bm}

\begin{document}


\title{Ultra high precision refractive index measurement of Si at $\gamma$-ray energies up to 2 MeV}


\author{M. M. G\"unther}
\email[]{m.guenther@gsi.de}
\affiliation{Helmholtz-Institut Jena, Fr\"obelstieg 3, 07743 Jena, Germany}
\affiliation{GSI-Helmholtzzentrum f\"ur Schwerionenforschung GmbH, Planckstrasse 1, 64291 Darmstadt, Germany}
\author{M. Jentschel}
\email[]{jentsch@ill.fr}
\affiliation{Institut Laue-Langevin, 71 Rue des Martyrs, 38000 Grenoble, France}
\author{A. J. Pollitt}
\affiliation{Institut Laue-Langevin, 71 Rue des Martyrs, 38000 Grenoble, France}
\author{P. G. Thirolf}
\affiliation{Ludwig-Maximilians-Universit\"at M\"unchen, Am Coulombwall 1, 85748 Garching, Germany}
\author{M. Zepf}
\affiliation{Helmholtz-Institut Jena, Fr\"obelstieg 3, 07743 Jena, Germany}
\affiliation{School of Mathematics and Physics, Queen's University Belfast, BT7 1NN, UK}


\date{\today}

\begin{abstract}
The refractive index of silicon at $\gamma$-ray energies from 181 - 1959 keV was investigated using the GAMS6 double crystal spectrometer and found to follow the predictions of the classical scattering model. This is  in contrast to earlier measurements on the GAMS5 spectrometer, which suggested a sign-change in the refractive index for photon energies above 500 keV.  We present a re-evaluation of the original data from 2011 as well as data from a 2013 campaign  in which we show that systematic errors due to diffraction effects of the prism can explain the earlier data. \end{abstract}

\pacs{41.50.+h, 07.85.-m, 07.85.Fv, 07.85.Jy, 07.85.Nc, 42.50.Xa, 78.20.-e, 12.20.-m}

\maketitle

\section{Introduction}
The rapid evolution of bright X-ray sources makes the knowledge of optical properties of  materials increasingly important (e.g. ELINP\cite{ELINP}, XFEL\cite{XFEL}) from the point of view of novel techniques and applications. In the few keV regime, extremely small foci and polarisers with unprecedented purity are revolutionising measurement techniques. At the same time optical properties provide a sensitive observable allowing fundamental interaction processes of waves with matter to be probed. The dispersion and refraction of electro-magnetic waves have been theoretically described by R. Kronig and H. A. Kramers \cite{deL.Kronig:26, Kronig1928}. The wavelength dependent forward atomic scattering amplitude can be written as a complex scattering amplitude $f(\omega)=f_0 +f'(\omega)+if''(\omega)$, where $f_0$ is the frequency independent part, while the complex term describes the frequency dependent part of the scattering amplitude, which is known as the anomalous scattering factor \cite{PhysRevA.35.3381, Gold64}. Via the Lorentz relation 
\begin{equation}
n=1+\frac{r_e \lambda^2}{2\pi}\Sigma_k(N_k (Z_k+f'_k+if'_k))
\end{equation}
the scattering properties of an atom are related to the macroscopic description of the index of refraction \cite{PhysRevB.30.640, Gold64}, where $r_e$ is the classical electron radius and $\lambda$ is the wavelength of the radiation. The sum $k$ covers all $N$ atoms of the sample with the atomic charge number $Z$ and the anomalous scattering factor of each $k$-th atom. Therefore the forward scattering amplitude can be accessed directly by measuring the refractive index and insight can therefore be gained into  the frequency dependence of scattering processes of the irradiated material. In the X-ray energy range, refractive index measurements were performed with the main motivation to investigate the forward atomic scattering amplitude \cite{PhysRevB.30.640, PhysRevB.42.1248}. From an application point of view, the knowledge of the X-ray refractive index was important to realise first refractive optics, which are well established in today's X-ray applications in biomedical, physical and material sciences \cite{Snigirev, 0022-3727-38-10A-042}.
To date, the refractive index of some materials has been experimentally determined mainly up to a photon energy of 133 keV \cite{PhysRevA.91.033803}. Materials with a low atomic charge number $Z$ were primarily investigated, since photo-absorption increases strongly with $Z$. As a first approximation for the calculation of the dispersion curve, the real part of the complex forward scattering amplitude in a classical non-relativistic approximation can be assumed to be dominated by Rayleigh scattering. Using the Thomas-Reiche-Kuhn \cite{Thomas1925, Reiche1925, Kuhn1925} sum rule $f'_0=-Zr_e$, the real part of the index of refraction can be written as
\begin{equation} \label{e1}
\delta(E_\gamma)=-Zr_e 2\pi \frac{(\hbar c)^2 N_A \rho}{E_{\gamma}^2 A} \mbox{.}
\end{equation} 
Here $c$ is the vacuum speed of light, $N_A$ is Avogadro's constant, $E_\gamma$ the $\gamma$-photon energy and $A$ the atomic mass. This classical approximation for the Rayleigh scattering is well established in atomic scattering processes in the low X-ray energy range and in the case of light atoms with low $Z$. 

\begin{figure*}
\includegraphics[width=1.\textwidth]{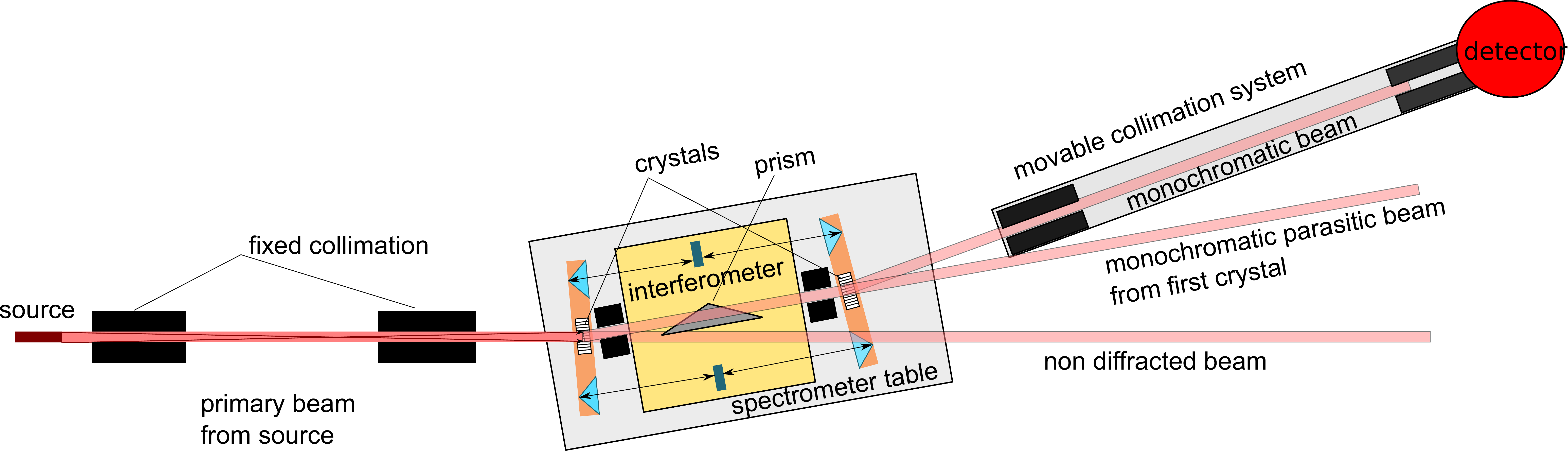}
\caption{\label{fig0}(color online) General layout and working principle of a double crystal spectrometer as used for the refractive index measurement. The photon source is 15 - 20 meters away from the first crystal inside the high flux reactor of the Institut Laue Langevin. Over this distance the beam is shaped by a system of fixed collimators. The spectrometer generates a low divergence monochromatic  beam, which is separated from the primary beam by a system of movable collimators. }
\end{figure*}

Information on the refractive index at much higher energies - in the MeV $\gamma$-energy range - was only accessible via extrapolation of the classical model. However, since the underlying interactions and scattering physics change in relative importance with increasing photon energies, the question arises up to which energy the extrapolation remains valid. The development of highly brilliant tuneable $\gamma$-ray sources further motivates this research \cite{ELINP}. Certain applications envisaged on these facilities, such as  nuclear resonance fluorescence (NRF) based detection, radiography and transmutation experiments, would benefit significantly from focusing optics in the MeV regime. Therefore, from a fundamental, but also applied physics point of view, it is important to establish a reliable experimental knowledge of the refractive behaviour of electromagnetic radiation at $\gamma$-ray energies. 

A first experiment to measure the index of refraction of silicon in the $\gamma$-energy range from 181 keV to 2000 keV was performed in 2011, using the high resolution double flat crystal spectrometer GAMS5 at the ILL in Grenoble \cite{PhysRevLett.108.184802}. The index of refraction is typically denoted as a complex number $n(\omega)=1+\delta(\omega)+i\beta(\omega)$, where the real part $\delta$ describes the phase shift of the electro-magnetic wave after propagation through matter, while the imaginary part describes the absorption. In the X-ray energy range, the decrement $\delta$ from $n=1$ is tiny ($10^{-5}$ to $10^{-7}$) and has a negative sign \cite{AlsNielsen2011}. However, in the 2011 experiment, an unexpected change in the sign of $\delta$ from negative to positive was observed above 500 keV and efforts at interpreting the underlying physics were made. A first attempt in reference \cite{PhysRevLett.108.184802} to attribute the sign-change to virtual pair creation, like Delbr\"uck scattering, turned out to be inadequate to explain the experimental findings, as detailed theoretical work indicates that the contribution from Delbr\"uck scattering to the real part of the refraction index is many orders of magnitude too low to account for the observed effect \cite{PhysRevLett.110.129501,PhysLettA.380.3703}. To further intensify the search for the underlying physics, a campaign was launched to look into the scaling of the refractive index with atomic number (Ge) in 2013. The possibility that systematic errors may have affected the result was also pursued, with a focus on temporal drifts in the spectrometer and re-measuring the identical silicon prisms from 2011. The 2013 experiment also employed with a slightly different measurement sequence (allowing better drift correction) and also better statistics. The 2013 campaign showed the same sign-change effect again for silicon, but it was not observed in a germanium prism, motivating further detailed studies. However,  a direct continuation of this activity was not possible, since the GAMS5 was decommissioned at the end of 2013 and until 2015 no further experiments were possible.
Interest renewed in the second half of 2015, when two new instruments, GAMS6 and DIGRA, were again put into operation, allowing for a continuation of the research activity. The research was taken up by a new cooperation between the Helmholtz Institut Jena (Jena, Germany), the ILL (Grenoble, France) and the LMU (Munich, Germany), with the aim of measuring a wider range of elements to elucidate the underlying physics and with a view to developing gamma ray optics. The campaign also had the expressed aim of eliminating the possibility of any diffractive effects affecting the measurement by including materials in the liquid phase (these results will be discussed in a forthcoming publication). The current publication focuses on a re-evaluation of the 2011 data together with the 2013 experiment on GAMS5 in the light of the recent results on GAMS6. In this publication we focus  on the silicon data. New data from a second set of silicon prisms is presented as well as a dedicated investigation of the refractive prism used during 2011 and 2013 experiments. Diffraction in the 2011/13 prism set is identified as the systematic error that affected the earlier measurements and gave rise to an (erroneous) interpretation of a positive refractive index decrement $\delta$. Furthermore, we report on a new measurement of the refractive index on GAMS6, using an improved setup and a new silicon prism set, which eliminates many potential systematic errors identified in the first generation experiments on GAMS5. The new findings show that the dispersive behaviour of silicon at $\gamma$-ray energies is in agreement with existing theory.

\section{General experimental methodology}
\subsection{Concept of the experiment}
The general concept of our refraction index measurements is well known and based on prism optics \cite{hecht2012}. In the visible range of the electro-magnetic spectrum this method has been established for more than a century \cite{Abbe1874} and later applied in the soft X-ray regime \cite{PhysRevB.30.640}. The principle consists in i) defining a low-divergence monochromatic photon beam, ii)  deflecting it via refraction at the interfaces of a prism and iii) measuring the deviation angle with respect to the incident beam.
The experimental setup has to fulfill several requirements: collimating the incident beam, monochromatizing it and analyzing the direction at the output. According to the classical model, the refractive effect decreases strongly with photon energy ($\sim -1/E^2$) within the X-ray or $\gamma$-energy regime. Therefore the experiment should increase its sensitivity with increasing energy. In our experiment, the function of primary beam monochromator and collimator is fulfilled by the first single crystal. Photons coming from the source are diffracted within a particular energy band width $E_\gamma \pm \Delta E_\gamma$ and within a certain angular range $\theta_B \pm \Delta \theta_B$. The angular width $\Delta \theta_B$ is the so-called Darwin width \cite{AlsNielsen2011} and $\Delta E_\gamma = E_\gamma \cdot \Delta \theta_B / \theta_B$, where $\theta_B$ is the Bragg diffraction angle. After the first crystal the propagation angle of the monochromatic beam is analysed via a second single crystal. Inserting the refracting prism between two single crystals allows the deflection to be measured by rocking the second crystal, while keeping the first crystal fixed to diffract the same photon energy. The refractive index for this particular energy can be derived from the angular deflection due to the prism. This process can then be repeated for different energies to obtain the energy dependence of the refractive index. This general concept of the measurement was the same in all experiments, although the particular realisation was slightly different (see details in the according sections).
Both crystals prepare and analyse the beam via diffraction and the sensitivity of the experiment is directly related to this process. Via dynamical diffraction theory \cite{RevModPhys.36.681, Laue} it can be shown that  ${\Delta\theta_B}/{\theta_B} \simeq 10^{-6}$. This relation refers to the angular width $\Delta \theta_B$ of the diffracted beams. The sensitivity with respect to angular shifts $\delta \theta_B$  is typically two to three orders of magnitude smaller than the width and therefore we expect it to be in the order of $10^{-8}$ of the Bragg angle. From Bragg's law it can be easily shown that $\delta \theta_B \sim \Delta \theta_B \sim 1/E_\gamma$, while the refractive effect of the prisms scales with $\sim -1/E^2$. Therefore it is already evident, that such a measurement will be sensitive only up to some maximum energy. The sensitivity is further affected by additional errors in the angle measurement  (vibrations, goniometer drift error etc.). 

\subsection{Photon source and beam shaping}
For our experiments the $\gamma$-beam was provided from an in-pile target inserted in an irradiation position in the high-flux neutron research reactor at the Institut Laue Langevin (Grenoble, France). We used powder samples consisting of 10 grams of Gd$_2$O$_3$ or 6 grams of BaCl$_2$, respectively, held in graphite containers \cite{KesslerJr2001187}. These containers were placed in a beam tube close to the reactor core and irradiated by thermal neutrons with a flux of around $5 \times 10^{14}$ s$^{-1}$ cm$^{-2}$. Therefore, the $\gamma$ beam is produced by neutron capture nuclear reactions on the $^{155, 156}$Gd or $^{35}$Cl isotopes of the samples with subsequent prompt $\gamma$ emission. The $\gamma$-photon emission rate was up to 10$^{16}$ s$^{-1}$. The produced $\gamma$-rays are pre-collimated by a collimator system over a distance of 15 m for DIGRA, 17 m (for GAMS5) or 20 m for GAMS6 from the $\gamma$ source (the source size across the beam is $2 \times 20$ mm$^2$) to the spectrometer. The divergence of the beam is in the order of $10^{-4}$ rad in the horizontal plane (diffraction plane of the crystals) and $10^{-3}$ rad in the vertical plane. Behind the spectrometers and their diffracting crystal, there is a second movable collimation system of 3 meters length. It separates the diffracted from the direct beam. At the end of the movable collimator, a high purity germanium detector is mounted for counting, ensuring that only diffracted photons are counted. Using the energy resolution of the Ge detector, an energetic region of interest can be set. This allows further suppression of background gamma rays and higher Bragg diffraction orders to be excluded. 

\subsection{Crystal Spectrometers}
Data obtained with three distinct double crystal spectrometers is discussed in the this paper: DIGRA, GAMS5 and GAMS6. In the case of DIGRA, the crystals are rigidly mounted on commercial goniometers (MICOS PRS-200), providing an angular resolution in the order of $5 \times 10^{-6}$ radian. The goniometers are additionally mounted on XYZ translation tables, allowing crystals to be scanned in position. The  angular resolution of the instrument is limited (when compared to the GAMS instruments) and was used for diffraction measurements in the Si prism, where a combination of translation and rotation features was important.
For the actual index of refraction measurement, the diffraction angles of the crystals need to measured with extremely high angular resolution. For this purpose, on both instruments, GAMS5 and GAMS6, the crystals were rigidly mounted on a double stage rotary axis. The axis is driven by a first stage, consisting of a backlash-free mechanical rotary table (PI-M-048.00) for displacements down to $10^{-6}$ radian over a range of $\pi/6$ rad. The second stage is a home-made Piezo-Flexure drive for displacements down to $10^{-9}$ radian over a range of $10^{-5}$ rad. The angular position of the combined axis is controlled by optical angle interferometers, providing sub-nanoradian angular resolution. In the case of GAMS5, the rotation axis of each crystal was controlled by an individual heterodyne Michelson interferometer. These interferometers follow the optical layout described in \cite{KesslerJr2001187} and were operated in air. The interferometer scheme is made highly symmetric to minimize the effect of temporal variations of the refractive index of air and glass on the angle measurements. Additionally, all environmental parameters like temperature, air pressure and humidity were recorded and used to correct temporal drifts. In the case of GAMS6, a different type of interferometric measurement is used. Two Mach-Zehnder type interferometers, but based on common optical components, allow each axis to be measured individually as well as their relative position. The layout of the interferometer is made such that any drift of optical components and glass refractive index would introduce the same measurement error on both crystal axes. As a consequence, the total error cancels out and time dependent drifts are minimized. To eliminate the effect of changes in the refractive index of the ambient air due to moisture, temperature and pressure, the entire spectrometer is operated under vacuum.
The interferometer electronics allow for an online monitoring of axis vibrations, providing a measurement of the uncertainty for each angle position. For GAMS6 an improved collimation was also used. Rather than using lead as the material for the inter-crystal collimation, two pairs of polished tungsten carbide plates of $2 \times 10$ cm length were mounted on precision translation stages. This increased the collimation contrast substantially, particularly for high gamma ray energies.

\section{Refractive index measurement at GAMS5}
\subsection{GAMS5 Experiment setup}
\begin{figure}
\centering
\includegraphics[width=0.47\textwidth]{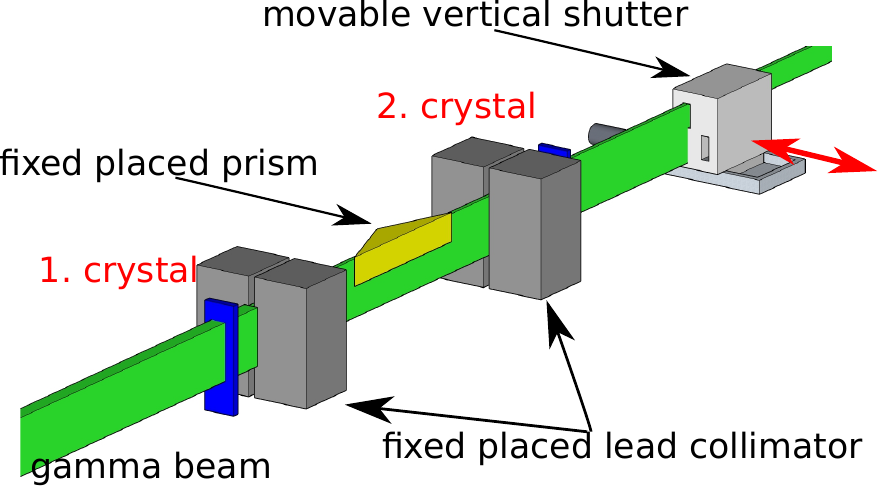}
\caption{\label{fig1}(color online) Experimental set up of the refractive index experiment at GAMS5. The green cuboid indicates the collimated $\gamma$ beam. The red arrow demonstrates the direction of the shutter movement to alternate between refracted and reference beam. }
\end{figure}

The experimental setup of our first campaigns in 2011 and 2013  to measure the refractive index of silicon is illustrated in Fig.~\ref{fig1}. 
An equilateral silicon prism with an angle of 160 degree and optically polished faces was  placed between the two silicon crystals of the spectrometer together with an additional collimation system made from two pairs  of 5 cm lead blocks  to ensure a spatially well defined  gamma flux. The prism was made such that it was covering only half of the 20 mm beam height. Switching between the upper and lower beam half allowed  $\gamma$-rays from a (non-refracted) reference and a refracted beam to be compared. The beam switching was realised using a movable height selector behind the spectrometer.

\begin{figure}
\includegraphics[width=0.48\textwidth]{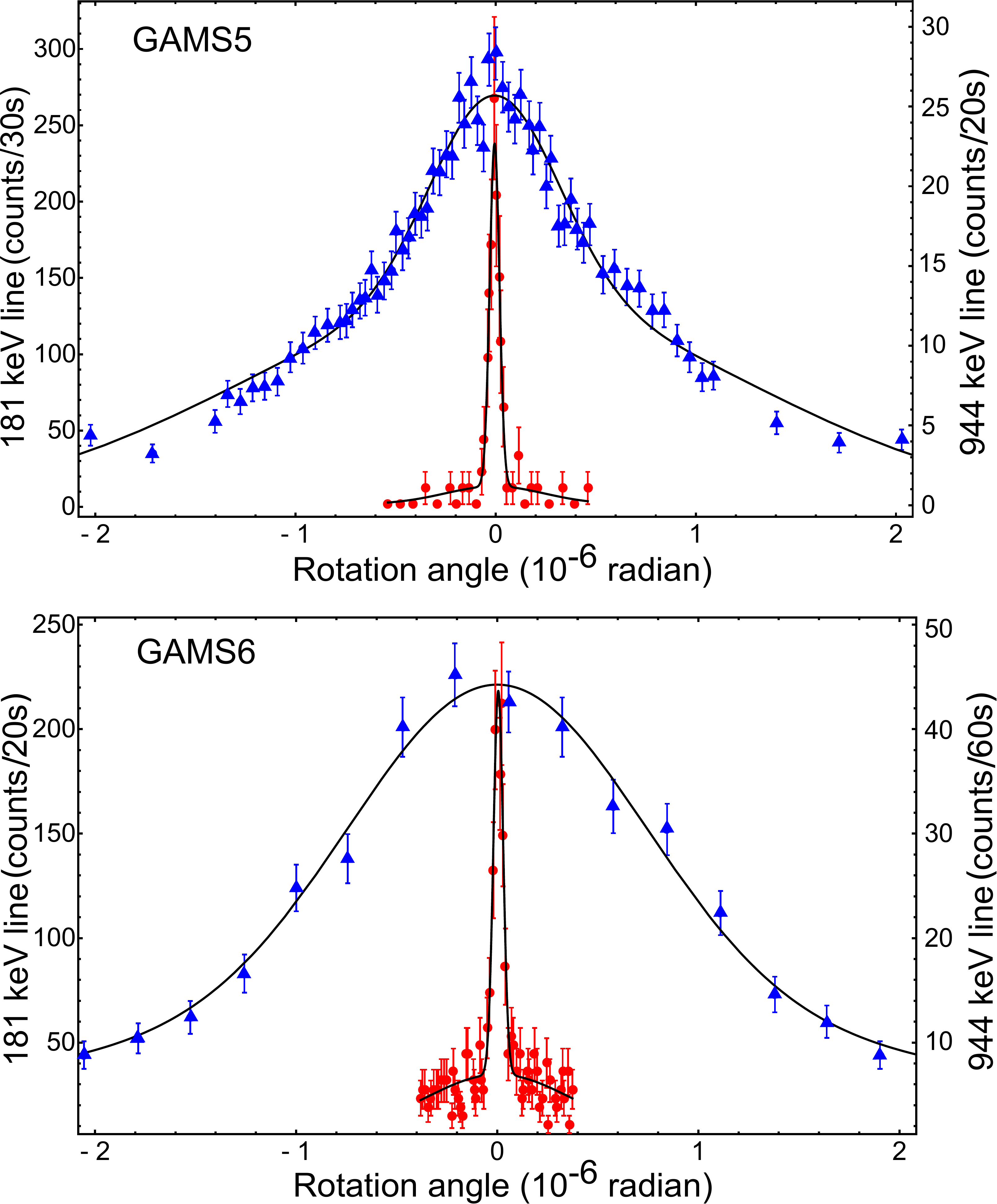}
\caption{\label{fig2}(color online) Intensity of 2 selected $\gamma$ lines as a function of the rotation angle of the second spectrometer crystal. The blue/red triangles/circles are the experimental data and the black line shows a double Gaussian fit to the data. The fit is used to determine the position of the center of the rocking curve. The rocking curves at 181 keV (blue triangles) and 944 keV (red circles) are shown as examples. For high $\gamma$ energies the width of the rocking curve becomes smaller with the $\gamma$ energy as $E_{\gamma}^{-1}$ in accordance with dynamic diffraction theory \cite{AlsNielsen2011}.}
\end{figure}

In the first experiment in 2011, the measurement sequence consisted of a simple alternate measurement of refracted and reference rocking curves by using the upper and lower halves of the beam, respectively. The angular scan direction of the rocking second crystal was the same for all scans. This measurement scheme results in so called "2-pack" data - pairs of two scans. During the analysis it turned out that the angular measurements were not completely free of temporal drifts (see discussion below). Therefore the 2013 experiment was made using an improved measurement sequence consisting of four scans: i) a positive scan of the reference beam ii,iii) a negative and a positive scan of the refracted beam and iv) a negative scan of the reference beam. The refraction angle was determined from a linear combination of this so called "4-pack" with the goal of reducing the impact of the linear temporal drift.

\subsection{Data analysis and results}
\begin{figure*}
\centering
\includegraphics[width=0.9\textwidth]{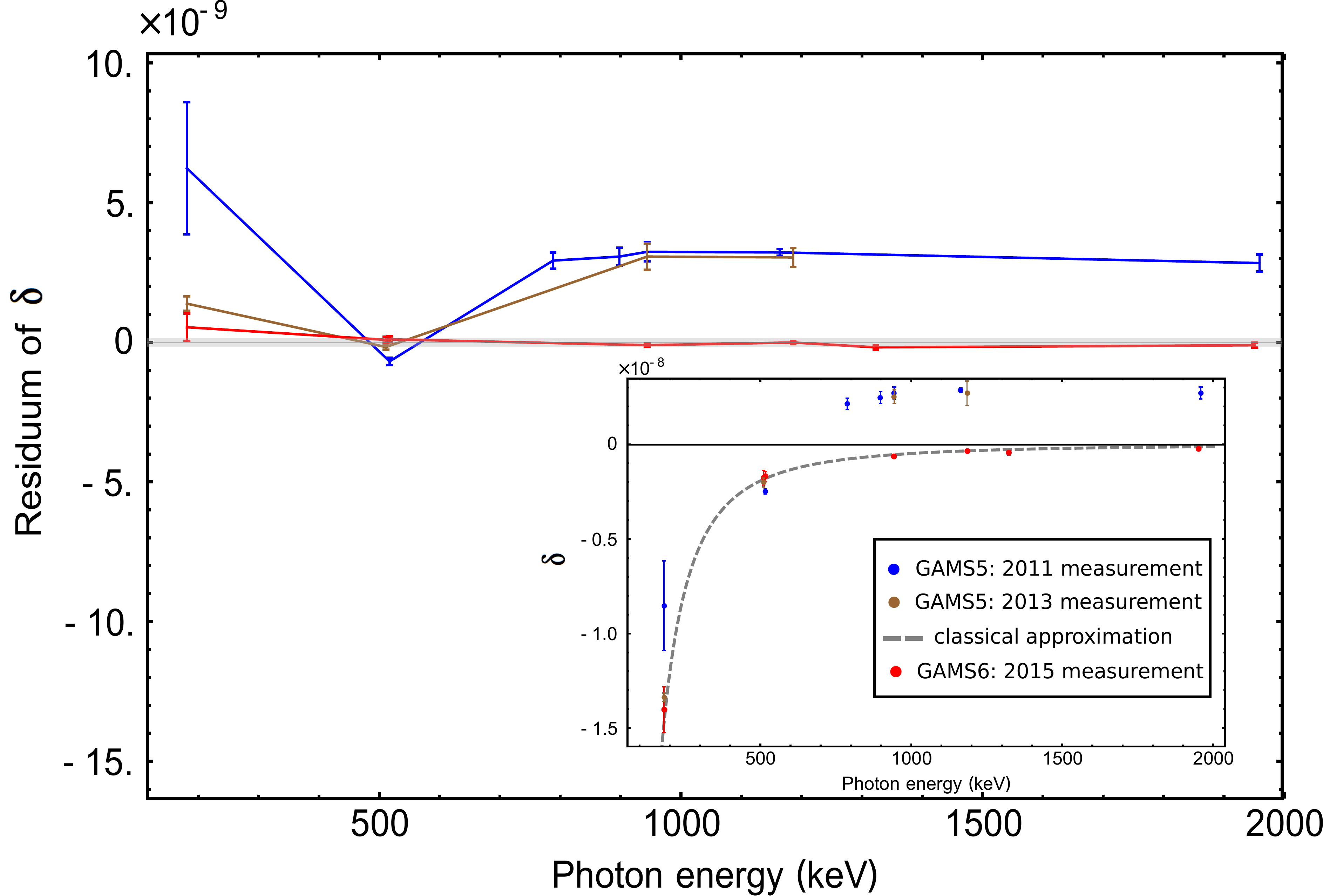}
\caption{\label{fig3}(color online) Results of the real part $\delta$ of the complex index of refraction as a function of $\gamma$-ray energy. The experimental data is plotted after subtraction of the classical theoretical model (\ref{e1}). The GAMS6 measurement follows the classical model well with decreasing uncertainties to higher energies due to the narrower rocking curve.
The GAMS5 measurements of 2011  and 2013 shows significant deviations at 181 keV and >500 keV. These deviations are caused by different systematic effects: at 181 keV they originate due to beam passage though different parts of the GAMS5 crystal and for energies >500 keV they arise due to the onset of diffracting phenomena of the silicon prism (for details see text). The grey bar indicates the sensitivity limit of the GAMS6 measurement, as established via a run without refracting prisms.}
\end{figure*}

The data analysis consisted of two steps: a) the fit of a theoretical lineshape to each scan to determine the peak position and b) the extraction of a difference of peak positions due to refraction. We used both lineshapes based on dynamical diffraction theory, as well as the sum of two Gaussians with the same peak position, but different width and intensity. Detailed comparisons of both approaches did not show any detectable difference in determining the peak position. For reasons  of simplicity the "two gaussian" approach was adopted. 
For the second step, the determination of the difference in peak position, a straight forward approach for the 2011 campaigns would be to use the difference of the two scans of the "2-pack" data. The results of this approach were published in \citep{PhysRevLett.108.184802}. To minimize the impact of temporal drifts, we also made a regrouping of  three "2-packs" into two "3-packs" and applied this approach to the 2011 data. The 2011 result presented below comes from this approach. For the 2013 campaign the difference of peak positions was obtained as linear combination of the four scans. The experimental value of the angular peak shift, according to the three possible evaluations, was extracted as follows:

\begin{eqnarray} \nonumber
r^{(i)}_2 &=& c^{(i)}_{u+}-c^{(i)}_{l+} \\ \nonumber
r^{(i)}_3 &=& 0.5 \cdot \lbrace c^{(i)}_{u+}-0.5\cdot(c^{(i-1)}_{l+}-c^{(i)}_{l-}) - \\ \nonumber
				&	&  c^{(i+1)}_{l+} + 0.5\cdot(c^{(i)}_{u+}-c^{(i+1)}_{u-}) \rbrace \\ \nonumber
r^{(i)}_4 &=& 0.5 \cdot (c^{(i)}_{u+}+c^{(i)}_{u-}-c^{(i)}_{l+}c^{(i)}_{l-})	
\end{eqnarray}

Here $c$ is the measured peak position, the indices $u/l$ indicate upper (refracted) and lower (reference) beam and $+/-$ the scan direction of the rocking crystal, the index $i$ is counting over the number of "packs". The errors of peak positions $\Delta c$ are obtained from an evaluation of the covariance matrix of the fit and are then propagated to yield $\Delta r_j^{(i)}, j=2,3,4$. Examples of individual scans fitted by a theoretical lineshape of the rocking curves are shown for 181 keV and 944 keV, respectively, in Fig.~\ref{fig2}. For each energy typically a few tens of 2-/4-packs were measured, yielding a corresponding set of values $ \lbrace r^{(i)}_j(E) \rbrace, j=2,3,4; i=1,...,N$. To extract a particular angular value $\langle r_j(E)\rangle$, two approaches were applied: i) a constant value was fitted into the set of data, the according error was extracted from the covariance matrix of the fit; ii) a weighted average was calculated and as error the standard deviation of the data was taken. It is worth mentioning that the original data of \cite{PhysRevLett.108.184802} was obtained following the first approach, while in the current paper we follow the second approach. The value of $\delta(E)$ can be obtained by solving the prism equation

\begin{eqnarray}
&&\langle r_j(E)\rangle = \\ \nonumber
=&&\left[\alpha-\arcsin\left\lbrace(n\cdot\sin(\theta_0)-\arcsin\left(\frac{\sin(\alpha)}{n}\right)\right\rbrace\right] \mbox{.}
\end{eqnarray}

The angle of $\alpha$ denotes an unknown offset angle between the incoming beam and the prism base line, while $\theta_0$ denotes the prism angle itself. The angle $\alpha$  is fitted to the data and is typically very small.
In Figure~\ref{fig3} a comparison of the silicon dispersion curves of the 2011 and 2013 campaigns with GAMS5 are shown. For better visibility of the results, we have subtracted the classical model (\ref{e1}) from the data and show only the residua. The measurements of the two campaigns clearly  agree with each other. Both experiments clearly show a very pronounced deviation from classical theory. The sign-change at an energy of $E>500$ keV is particularly noteworthy. A significant deviation at 181 keV is also observed, which can be  explained by the fact that the upper and lower beam were passing through different regions of the crystals. Here a slight mismatch of lattice spacing caused by a temperature gradient or mounting strain might introduce this effect. Such an effect is more strongly pronounced for lower energies \cite{Mana:ks5036}. This hypothesis was verified during the 2013 campaign. The experiment was repeated without any prism mounted between the crystals, which should cause a zero result. In this zero measurement  a slight angular offset was found for 181 keV. For all higher energies the result was consistent with zero.

\section{Radiography of the Silicon Prisms.}
\begin{figure}
\includegraphics[width=0.48\textwidth]{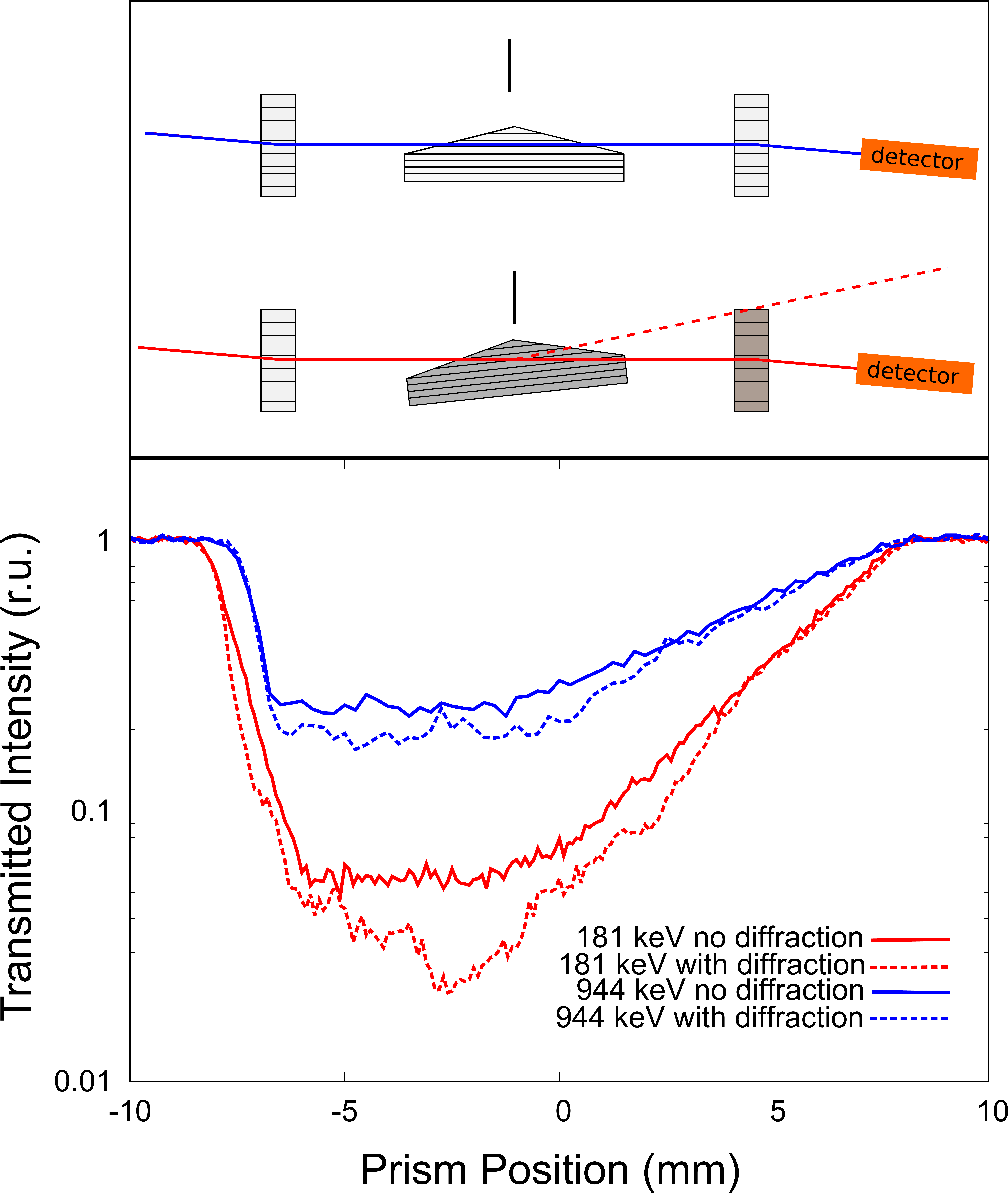}
\caption{\label{radiography1}(color online) Results of a radiography of the silicon prism as it was used for the GAMS5 experiments. The solid lines show the intensity profile, when the prism was not in diffraction orientation. The dashed lines show the same measurement, but with the prism in diffraction position. It can be seen that substantial amounts of intensity can be diffracted.}
\end{figure}

To address the deviation of the refractive index from the classical behaviour of equation (\ref{e1}) and the sign-change observed in silicon, an intensive experimental campaign was undertaken. A range of materials (particularly liquids) of differing atomic number Z as well as a new set of silicon prisms were investigated. A particular focus was put on the question, whether the silicon prisms showed diffraction effects at angles close to those used in the experimental arrangement and affect the refractive index measurement by adding a diffractive component to the angular deviation of the beam. The silicon prisms were investigated separately, using the double axis diffractometer DIGRA, as it allowed both linear translation and rotation of the prism through a monochromatic, microradian divergence beam. The general setup is shown in  Fig \ref{radiography1}. In a first measurement, the prism was rocked at several energies of the beam, while the detector measured the intensity of the transmitted beam. Assuming the prism material to be non-diffracting and a rather narrow rocking angle range (less than 2 degrees), no variation of intensity was expected. However, strong effects were detected and associated with diffraction of the $\langle200\rangle$ planes of the silicon material parallel to the base side of the prism. The effect was studied for energies from 181 keV up to 1 MeV and it could be shown that up to 50$\%$ of the intensity could be diffracted. The angular acceptance of the diffraction (FWHM of the rocking curve) was measured to be in the order of 500 $\mu$rad. This is much larger than the angular acceptance of a perfect crystal and can be possibly explained due to strain from the surface polishing process. Based on this measurement, it was possible to reconstruct a scenario explaining the angular deviation measured in the 2011 and 2013 experiments as a superposition of both refraction and diffraction, rather than an anomalous refraction effect with positive decrement $\delta$. We constructed a simple model, based on the refractive deviation of the beam scaling as $\sim -1/E_\gamma^2$, while the diffractive part scales as $\sim 1/E_\gamma$ with photon energy. Further it was assumed that the diffractive part only starts to superpose the refracted beam  above a certain energy. To understand this assumption, it is important to notice that we assume the diffraction to be caused by a fraction of the primary beam, which has not been diffracted by the first crystal of GAMS5. This beam has also a divergence of the order of 500 $\mu$rad, which is matched by the acceptance of the prism. In order to enable diffraction from the prism, the primary beam has to be within the acceptance range for diffraction. Since the angle of the incidence of the primary beam onto the prism changed with increasing energy, there will be a critical energy from which on diffraction takes place.  In order to fit the experimental data of the 2011 GAMS5 experiment, we defined therefore the following model:

\begin{equation} \label{erefdif}
m(E)=-C/E^2+\frac{1}{1+e^{-(E-E_c)/\delta E}}\cdot D/E
\end{equation}

\begin{figure}
\includegraphics[width=0.48\textwidth]{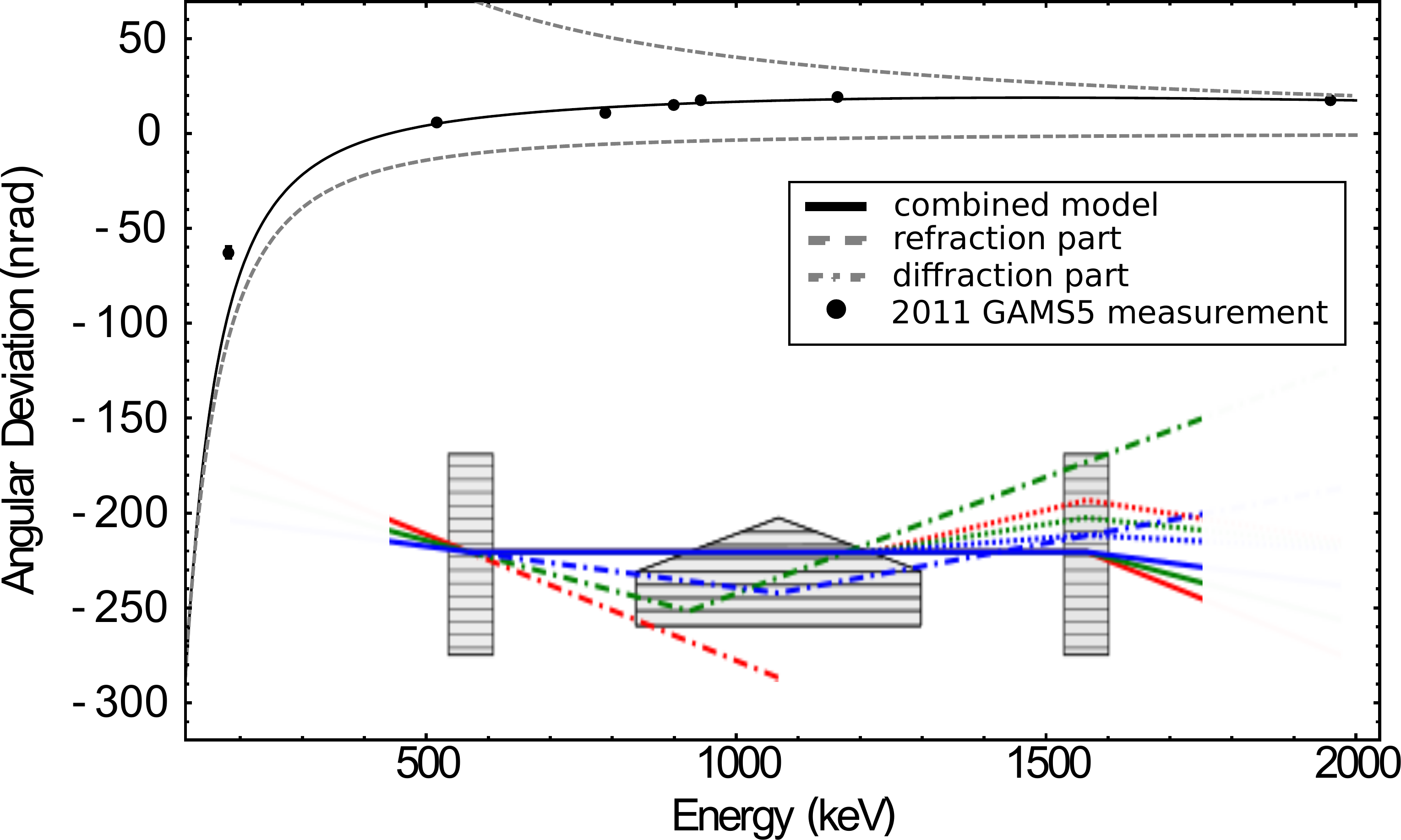}
\caption{\label{refraction_diffraction}(color online) Data from the 2011 GAMS5 run compared to a model (\ref{erefdif}) describing the measured angular deviation in terms of superposition of a refraction and a diffraction component. The parameters of the model were fitted to the data showing an excellent agreement: $D=0.000040661148613035$, $E_c=978.78$, $\delta E=446.61$. The refraction part in (\ref{erefdif}) was not fitted, but calculated using the classical approximation model (\ref{e1}):  $C=0.004146748006128347$. However, the data point at 181 keV is out of the combined model, as well as the refraction part, where the reason was explained in chapter III.B. . As shown in the insert, the primary beam is diffracted by the silicon lattice of the prism above a certain threshold  energy. As illustrated, the incident, low-divergence beam enters the angular acceptance range for  Bragg diffraction for higher energies. Note that the angles are exaggerated for clarity in the schematic and therefore  the diffracted and refracted beam appear to separate spatially. However, for the small angles encountered at high photon energies, both possible paths through the prism overlap and can enter the detection system.  }
\end{figure}

As the exact orientation of the prism in the 2011 and 2013 runs is not known, the combination of diffraction and refraction effects is described by scaling constants $C,D$, which correspond to the refraction and diffraction angles, while the parameters $E_c,\delta E$ describe the onset of diffraction. All parameters were fitted, yielding an almost perfect agreement of the model with the experimental data. This is illustrated in Figure \ref{refraction_diffraction}, which shows that the measured sign-change can be completely explained by the crystalline properties of the prism material. The combination of diffraction and refraction is therefore the underlying systematic error affecting the 2011 and 2013 GAMS5 measurements in silicon. In combination with our most recent results, which show agreement with existing theory using a different set of silicon prisms, it is evident that the report of anomalous refractive behaviour in silicon described in \cite{PhysRevLett.108.184802} has been superseded by our current results.
The new silicon prisms for the GAMS6 experiment were carefully investigated using the same approach. These prisms, although also made from crystalline silicon, were oriented in such a way that they did not show any pronounced diffraction phenomena potentially perturbing a refractive index measurement.

\section{Refractive index measurement at GAMS6}
\subsection{Experiment setup and systematics}
\begin{figure}
\centering
\includegraphics[width=0.48\textwidth]{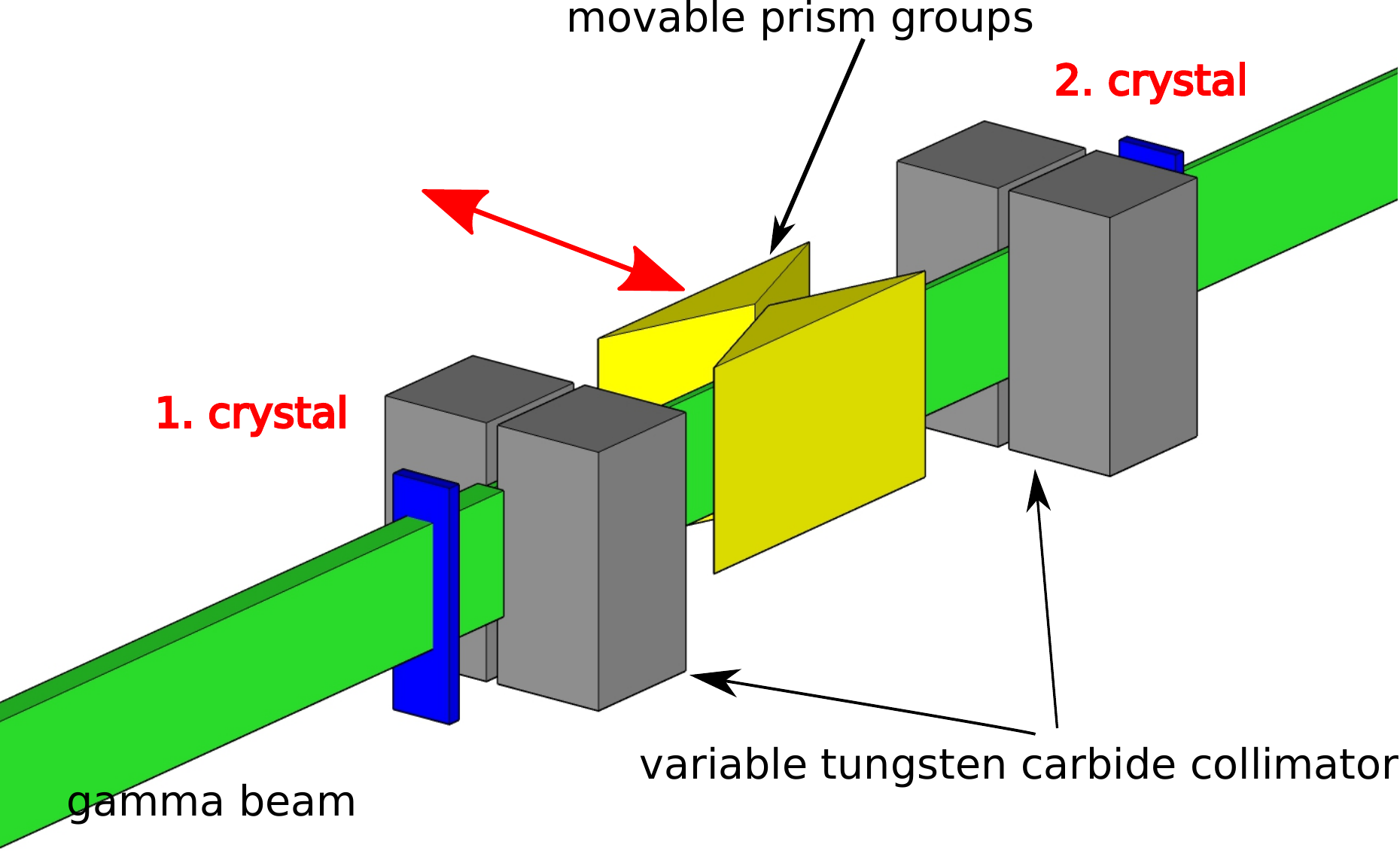}
\caption{\label{fig4}(color online) Schematical set up of the GAMS6 experiment, which illustrates the methodology of the experiment. Two prisms are shifted between the two crystals. This arrangement is less prone to systematic errors by ensuring that the same sections of the spectrometer crystals are used during the experiment and that refraction is measured in both positive and negative angles.}
\end{figure}

The main difference of GAMS6 with respect to GAMS5 is that it operates under vacuum and that the angular interferometers follow a completely new layout. The consequence is a substantially better temporal stability, allowing for more reproducible long-time measurements. The principle set-up is shown in Fig.~\ref{fig4}.

Although the main principle, based on Laue-Laue diffraction on two flat silicon crystals, is the same that was used at GAMS5, the layout of the refractive prisms is completely different. Rather than comparing a refracted and non-refracted beam, we decided to double the refraction effect by using oppositely aligned prisms of the same material and geometry. This allowed to the full height of the beam to be used, doubling the counting rate and hence improving the measurement statistics. Furthermore, this eliminates a potential systematic error of the GAMS5 experiment: the diffraction occurs always by the same crystal volume and both orientations contribute with the same statistical significance and with the same sensitivity to drifts to the final result. In order to switch between the two prism orientations, they were actively moved forward/backward within the inter-crystal collimator by a motorized precision translation stage. This became possible, since the entire inter-collimation system of GAMS6 is mechanically isolated from the optical interferometers and therefore the movements did not impact the angle measurements through vibrations. We used two pairs of equilateral prisms with an prism angle of 120 degree. They were mounted in a precision machined prism holder, enabling an exact and reproducible alignment of the prism base. A potential disadvantage of using an active displacement within the vacuum chamber would be a temperature encoding due to the motor. This point was carefully monitored during the experiments, using six thermistor probes distributed over sensible positions of the interferometer and on the crystals. Although the step motor itself showed some signature of heating (about 0.1 K variation), no relevant temperature variation (measurement sensitivity was 5 mK) on the crystals and on the interferometer was measured.

\subsection{Experiment results}
\begin{figure}
\centering
\includegraphics[width=0.48\textwidth]{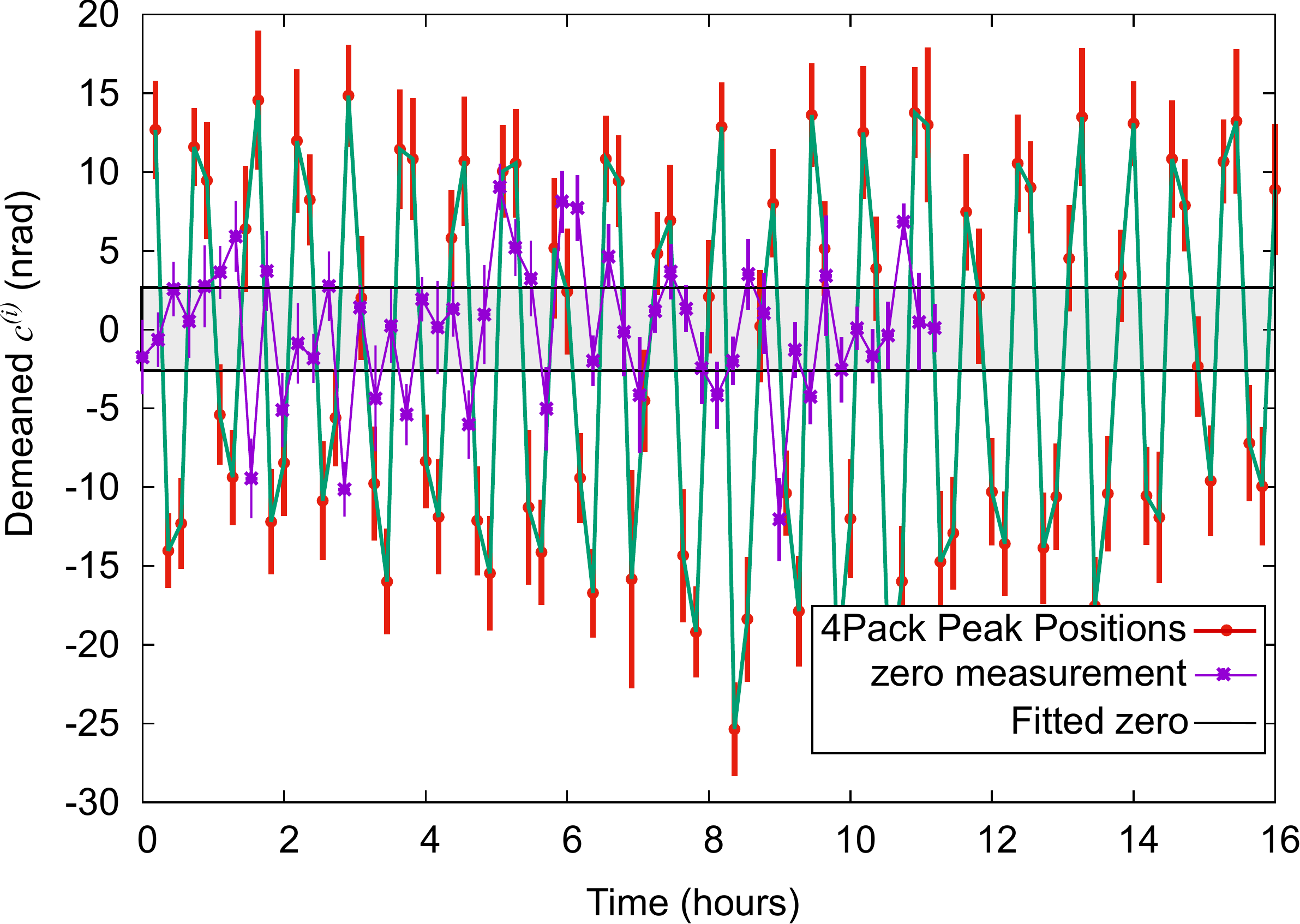}
\caption{\label{fig9a}(color online) Red error bars: Time sequence of measured peak positions $c^{(i)}_{u,l,+,-}$ for the 517 keV measurement. The pattern of the 4-pack sequence can be clearly identified. From the data a slight linear drift of -0.6 nrad/h has been subtracted to visualize the corrective functioning of the 4-pack algorithm. Violet error bars: These are data from a measurement sequence without prisms. The data scatter randomly. The black lines indicate the sensitivity after 12 hours measurement.}
\end{figure}

During the refractive index measurement, the oppositely oriented prisms were moved into the collimated beam and a rocking curve was measured in each prism position. Thus having two prism positions ($u,l$) and two scanning directions ($+,-$), a 4-pack algorithm was used to minimize eventual drift errors. The evaluation of the measured rocking curves was performed using a double Gaussian fit model. In order to obtain a quantitative estimate of the sensitivity of the experiment, we carried out a measurement series without prisms in the beam. This first order drift is in the evaluation procedure eliminated by the 4-pack algorithm. For illustration purposes we show in Figure \ref{fig9a} a time sequence of the fitted 4-pack peak positions $\lbrace c_{u+}^{(i)}, c_{l-}^{(i)},c_{l+}^{(i)},c_{u-}^{(i)}\rbrace$ from the measurement at the 517 keV on GAMS6 together with the measurement without prism. In order to illustrate the removal of linear drift, we subtracted a slight linear drift of -0.6 nrad/h from the data. During the real data processing the subtraction was not performed, since it is implicitly included by the 4-pack algorithm. 
The results of $\delta(E)$ after subtraction of the classical model (\ref{e1}) are included into Fig.~\ref{fig3}. The experimentally determined sensitivity, as determined from the measurement without prisms, is shown as a grey bar. The good consistency of the residua with zero demonstrates that the recently measured data fit the classical model well. 

\section{Conclusion}

The recent measurements of the refractive index of silicon using the GAMS6 spectrometer and a detailed investigation of diffractive effects of the original prisms using DIGRA  have allowed the anomalous GAMS5 results to be re-interpreted. Since the original publication \cite{PhysRevLett.108.184802}, a number of systematic errors have been identified, which allow the GAMS5 results to be explained. The largest contribution comes from diffractive effects, due to the crystalline nature of the silicon prism. Further systematic errors were found to arise from the fact that different sections of the diffracting crystals were used for different parts of the measurement (affecting the measurement at low energies). These factors have clearly been improved in the GAMS6 setup, resulting in substantially smaller error bars in Figure \ref{fig3}. From the experimental data obtained GAMS6, we conclude that the extrapolation of equation (\ref{e1}) holds up to energies of about 2 MeV in silicon. The experimental technique developed and applied with the GAMS6 setup has shown to be capable of delivering high quality data, allowing the refractive index of materials with MeV photon energies to be measured.


\end{document}